# Applications of Voice User Interfaces in Clinical Settings

**Akiri Surely**
University of Maryland
Baltimore County (UMBC)
Baltimore, U.S.A
sakiri1@umbc.edu

**ABSTRACT**
The adoption of modern technologies for use in healthcare has become an inevitable change. The emergence of artificial intelligence drives this digital disruption. Artificial intelligence has augmented machine capabilities to act like and interact with human beings. As the healthcare industry adopts technology in most areas, an area in healthcare that is touched by this change is clinical practice. New technologies are being designed to improve healthcare services. One aspect of these technologies is voice user interfaces. This paper reviews applications of voice user interfaces in clinical settings.

Several information sources were consulted, and based on eligibility criteria, a search was conducted, and ten papers selected. This study presents findings from the last ten years (2009-2019). The results are categorized based on findings, also they contribute to the discussion and the research gaps identified for future study as regards context-aware voice user interfaces and the appearance of conversational agents from a given set of options.

**INTRODUCTION**
In 1952, what was formerly known as Bell Laboratories, designed the first kind of voice technology. Audrey, a system that was able to recognize a specific voice speaking digits out loud. IBM followed shortly after that with Shoebox, a voice recognition system that understood and could respond to 16 words in the English Language. By the 1960's technology that could support nine consonants and four vowels existed. Advances in computing power, artificial intelligence, machine learning, and natural language processing saw Apple launching Siri in 2011, an intelligent assistant that relies on natural language processing and speech recognition. Other prominent examples of conversational agents include Alexa by Amazon, Cortana by Microsoft and Google Home. This explosion and advances have seen to it that technology to support voice recognition becomes inexpensive and powerful upholding Moore's Law. Thus, leading to the adoption of voice interfaces. Voice user interfaces are defined as what a person interacts with when they communicate with a spoken language application [10]. These interfaces can be applied in a variety of domain such as automobiles, home settings (television remote controls [29], washing machines, smart speakers, microwave ovens), military settings (Command and Control on the Move (C20TM), the Soldier's Computer, combat team tactical training and voice control of radios and other auxiliary systems in Army helicopters [32], wearable devices, mobile devices and healthcare settings among others. In any application where it is used, voice user interfaces most times undergo a set of generic steps when used. They include 1. Activation; 2. Automatic speech recognition; 3. Natural language understanding; 4. Action; 5. Natural Language generation. Most times the speech recognition module is the most arduous, as it has the task of filtering out noise and capturing the users' command accurately. Natural language processing allows humans to interact with machines using natural language. As technology proliferates around us, it has become increasingly important that human machine-interaction is no longer restricted to a few phrases as a few decades back but can mimic a natural conversation between humans.

Present findings from 2010 to 2017 that suggest that an increasing amount of Human-Computer Interaction (HCI) research that is focused on healthcare has been published in the annual Human Factors and Ergonomics Society (HFES) proceedings [30]. The healthcare landscape is changing exponentially, and certain technologies are responsible for far-reaching implications in terms of diagnostics, treatments, and delivery of care in the future [25, 33]. The most prominent being artificial intelligence. Artificial intelligence has been classified into four categories [18], 1. systems that think like humans; 2. systems that act like humans; 3. systems that think rationally; 4. systems that act rationally. The leap towards artificial intelligence has caused conversational agents to go from agents that allowed only constrained user input to unconstrained natural language input, complex dialogue management and overall flexibility in conversation. This constant improvements to conversational agents have opened opportunities for potential applications that play crucial roles in healthcare [14, 6] for all stakeholders involved. These interfaces are capable of helping clinicians with decision making, clinical documentation, improving patient outcomes, assisting persons with disabilities or elderly, as therapy assistants, presenting information about a disease.

This paper review ten works of literature discussing different voice user interface technologies that have been applied in clinical settings. These technologies have been applied in different domains in the clinical setting, Arash, social robot to support children with cancer [22], a cognitive coach for dependent persons [27], SimSensei Kiosk, a virtual human interviewer that can engage in face-to-face interactions with the user, while the user talks and shares information [11].

**BACKGROUND AND MOTIVATION**

The healthcare industry has come a long way, visualizing a few decades back to the present. In times gone by, hospital settings used slow desktop computers, ancient medical devices, landlines, fax and copy machines that minimized the speed of clinical workflows. However, today it is probable to see wireless communication, real-time locating devices, self;-service kiosks, remote monitoring tools, portal technology, telehealth, voice communication systems, and robotoids amongst others. According to Cassano [9], technology is the foundation of the future in healthcare and technological development in clinical applications is a trend that is not going anywhere anytime soon. This fact is largely attributed to consumer behavior. Healthcare stakeholders to maintain a good health status or generate better health outcomes.

As previously mentioned, artificial intelligence can be found at the heart of the current technological trend in healthcare. The potential for artificial intelligence in healthcare is broad, as artificial intelligence keeps emulating human behaviors precisely and at a lower cost. The earliest known application of artificial intelligence in healthcare dates back to 1964, the Dendral project. An expert system to help organic chemists identify unknown organic molecules [18]. There has since been a plethora of research, papers, and applications of artificial intelligence in healthcare. In medical diagnosis [28, 16, 2], in mental illness for detecting symptoms and in identifying suicidal patients [15], in performing solo surgical operations [20], IBM Watson's ability to detect treatments for cancer patients, Google cloud's healthcare application that helps organizations to manipulate data. Figure 1 shows the rising revenue from artificial intelligence startups in healthcare in the United States. It is shown that from 2013 to 2018 there has been a steady increase in the dollar amount (patients, physicians, employers, insurance companies, pharmaceutical firms, and government) are requiring more and demanding more in treatment and the world striving at large for value-based healthcare, there is a substantial need to innovatively manage health and lifestyles. Health technology innovations are changing the way things are viewed and how work is done together in the clinical world. A majority of the solutions technology has provided are in response to a consumer's need and desire to use technology generated from these startups.

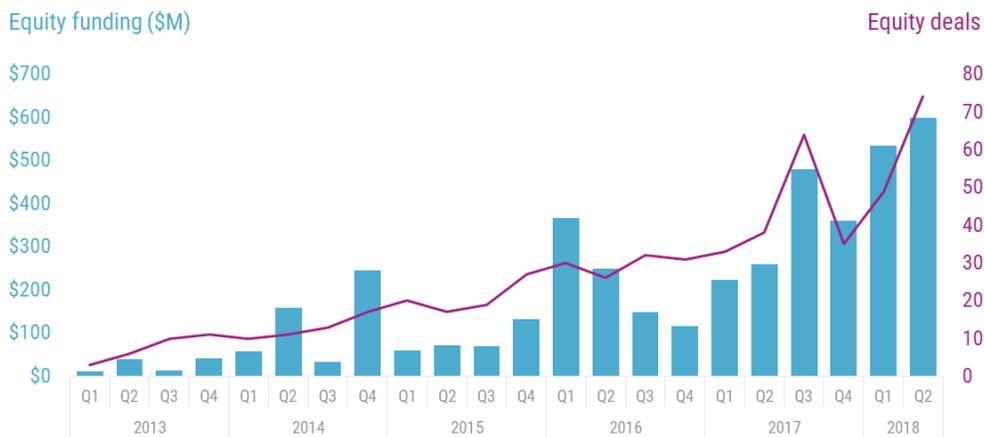

**Figure 1:** A bar chart showing the dollar amount healthcare AI startups have raised across 576 deals since 2013 in the United States. Retrieved from https://www.cbinsights.com/research/report/ai-trends-healthcare/

These benefits provided by artificial intelligence have sparked further interest in the use of voice user interfaces. Voice interfaces have become ubiquitous, embedded in our daily lives. These voice user interface technologies are often referred to in pieces of literature using different terms that can connote different meanings occasionally. They can be referred to as conversational agents, or intelligent or personal assistants, or embodied conversational agents, or speech recognition systems. Porcheron et al [24], adopts the term conversational interfaces, technologies that enable users to "have a conversation" with and "just ask" questions of. The pie chart depicted in figure 2, shows 47.1 percent of voice technology startups across all verticals that are focused on a single sector are focused on healthcare as cited by Brownstein et al. [8]. These statistics obtained encourage the study of voice user interfaces being adopted in healthcare.

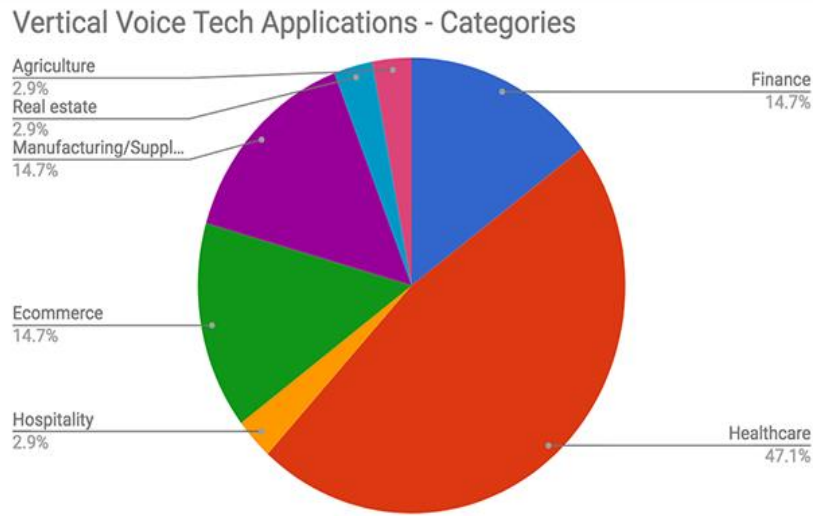

**Figure 2:** A pie chart showing vertical voice technology applications across voice technology startups. Retrieved from https://www.mobihealthnews.com/content/37-startups-building-voice-applications-healthcare

**LITERATURE REVIEW**

This is a systematic review of ten peer-reviewed papers chosen according to eligibility criteria, organized and analyzed to then identify essential gaps that can be tackled at a later time. The goal of this study is to synthesize findings from this literature into a narrative that discusses the main issues and gaps identified and possible future directions for voice user interfaces applied in clinical settings.

**Methods**

*Information sources*

For this review, a variety of sources have been used, namely journals, literature databases, and conference proceedings. The following list of journals was consulted: Journal of Medical Artificial Intelligence, The Journal of Presence Teleoperators & Virtual Environments, Journal of Artificial Intelligence for Medical Sciences, Journal of Intelligent and Robotic Systems, Journal of Biomedical Informatics, Journal of Engineering in Medicine, The International Journal for Speech Technology, SLAS Technology Journal.

The IEEE digital library was sought, a research database that grants access to journal articles and conference proceedings. Also, PubMed, a search engine that relies solely on MEDLINE database was used, Association for Computing Machinery (ACM) Digital Library (computer science), ACM SIGCHI, ScienceDirect, Elsevier, Springers and Taylor and Francis. Further, conference proceedings from the SIGCHI conference were searched from the years 2009 to 2019, International Conference on Autonomous Agents and Multi-agent Systems, International Conference on Social Robotics.

*Eligibility criteria*

This review is limited by three constraints: time, settings and interface. To begin with, technology is ever-changing and improving due to the rapid evolution of technology. For instance, in six decades voice technology has gone from a few syllables to wide vocabularies. As such this study has been restricted to include papers from the last ten years (2009-2019). To ensure that only relevant technology is included in the review.

Also, all papers chosen take place in the clinical setting. For this review, a clinical setting could be the location of a hospital, specialist clinic, outpatient facility, rehabilitation or wellness centers. Further, as part of the requirements two of the papers reviewed must be proceedings from the Conference of Human Factors (CHI). However, after searching in accordance with the stipulated years only one was found.

Finally, the last constraint of this review is the interface. Excluded systems were constrained non-spoken dialogue interface, where the system did not respond using voice technology or communication was restricted to voice recognition alone.

*Search*

A search was performed in October 2019 and updated in November 2019. The process of query construction was

based on the author's previous experience on the subject. The generic search query was as follows:

> voice user interface OR conversational agents OR relational agents OR intelligent virtual agents OR virtual agents OR embodied conversational agent AND (clinical setting OR hospitals OR health facility OR (health OR healthcare))

Also, the reference lists of relevant articles were searched to explore further sources. Grey literature identified in databases (including dissertations, theses, and conference proceedings), were also included for screening.

*Study Selection Criteria*

Study selection was conducted by screening titles and abstracts based on eligibility criteria, for qualified papers the full texts were then assessed. The articles included are:

- Written in English
- Applied in clinical settings not just in healthcare
- Include a conversational agent applied in a domain (training, engagement or assistance among others).
- More than three pages or less in length
- Working papers and peer-reviewed papers
- Easily accessible on the web.

**Results**

*Description of conversational agents*

**Frame vs Finite**

Most spoken dialogue systems adopt either a frame dialogue management or the finite-state dialogue management. The dialogue manager governs the interaction style. The finite-based dialogue the system controls the interaction. This management is sometimes adopted because it is straightforward to encode and there exists a clear mapping of the interaction to model. However, this method of encoding affords limited flexibility of interaction, consider the virtual nurse in Bickmore et al. [5] as seen in figure 3, the inputs are constrained to the output options provided on the screen. The patient can only select from the options provided on the screen. This is improper for complex, as patients cannot fully express themselves and so the help provided to the patient is restricted. Another example is the embodied conversational agent used for the Epworth Sleepiness Scale interviews in Pierre et al. [23], the embodied conversational agent conducted interviews based on the Epworth Sleepiness Scale (ESS) ignoring anything else the interviewee said that is not a direct answer to the system's question. While the frame-based dialogue management works in an opposing manner the user can answer the questions placed by the system using a wide range of answers. The system determines what was entered and what must remain. A set of rules determines the next step, question and information presentation. Roberta [27], as shown in Figure 3, was designed using frame-based dialogue management. The system allowed users to tell stories as they exercised their speech and memory capabilities. SimSensei [11] was also designed using frame-based dialogue. The adoption of frame-based dialogue management avoids strict constraints unlike in the finite dialogue architecture whereas the finite dialogue management is easier to adopt.

**Delivery modality**

Conversational agents are supported by different technologies as shown in figure 3. Humanoid robots [27, 1,], three-dimensional animations [26, 7, 11, 23, 5], web browser [4], Tablet [5] and PC-based Virtual Reality (VR) systems [17]. Further, agents that were delivered virtually were delivered mostly using human animations, SimSensei Kiosk, a virtual human interviewer [11], a virtual nurse embodying a human being for hospital discharge [5]. Others such as Bott et al. [7], the conversational agent was delivered using an animated avatar of a puppy. McDonnell et al. [21], presented findings that abstractly depicted cartoon characters were often considered highly appealing, friendly and more pleasant than realistically looking characters. The agent presented here was used as an eCoach for prostate cancer patients. Although, the data from the focus group conducted showed that some users did not adopt the use of a cartoon character and felt it belittled the situation.

**Input and output modality**

For this review, only spoken dialogue conversational agents were considered. For input, the majority of the conversational agents identified made use of speech, with except the virtual nurse for hospital discharge by Bickmore et al. [5]. This system made use of a touchscreen for input. The patient was presented a set of presumed possible responses on the screen and allowed to select a possible option that was then sent to the nursing station. The other studies took spoken natural language for input (9 of the 10 studies evaluated), requiring speech recognition.

**Nonverbal behavior**

Most of the studies evaluated implemented nonverbal behaviors from the system and the users. Ninety percent of the studies reviewed, conversational agents implemented nonverbal behaviors, such as posture shifts, hand gestures like a wave or an open arm or smiles [5, 27, 22], Care Coach [7] responds to users touch and petting with bodily reaction or by sometimes appearing to cry or sleep. Beveridge, M., & Fox, J [4], however, did not implement any nonverbal cues for the clinical decision support system. Perhaps because, the system was built on the Homey project, an old dialogue system. Whilst in terms of the system recognizing non-verbal behaviors from the users, only forty percent were able to recognize nonverbal cues by implementing facial recognition, such as SimSensei Kiosk [11] is able to track the intensity of smile from users to detect depression, Arash

[22], is able to spot a face or identify a voice with the help of sensors (Kinect sensor), camera and a microphone array.

## DISCUSSION

### Main findings

Despite the increasing applications of conversational agents in healthcare, there exists much work to be done in the area and their use in clinical settings as defined by this study is relatively rare. Upon evaluation, it can be inferred that most of the conversational agents are not relational agents except SimSensei and the virtual nurse [11, 5]. Relational agents are computational artifacts that are designed to maintain long-term social-emotional relationships with the users. They are built to remember history and manage users' future expectations. The specialization of a conversational agent as a relational agent is particularly important in health. The human race is still very skeptical towards adopting intelligent technologies as regards health-related matters despite the challenges in healthcare they address. Relational agents possibly play a role in alleviating this concern, they are designed to establish trust with users and thus increasing the motivation and desire for the user to interact with the system. Some examples of relational behaviors include: addressing the users properly with their names, titles or roles, interacting with the users socially at the end and beginning of conversations, humor is used as required relevantly and empathy is expressed appropriately among others. While it doesn't need to be implemented in all cases, for example, a relationship between the system and user is not always necessary [23]. Considering the use case of the conversational agent used for Epworth Sleepiness Scale interviews it is not required that it be relational.

Similarly, only two conversational agents in the papers reviewed account for context [27, 22]. Roberta uses context-aware platforms in the dialog. So, the topics of discussion are managed according to context. This personalization ultimately leads to better interaction. Dourish [12, 13], describes conversations as an embodied phenomenon because their structure and order are gotten from how participants act in real-time and under the constraints of their immediate environments. Further explaining, that actions and settings are fundamentally intertwined and without context, there is no action. In the design of a conversational agent, context should be considered, implementing context enables the system at runtime to decide how to respond to the user based on the current context of use by the user. For example, conversational agents expressing empathy by employing of non-verbal cues such as a sad face or low demeanor is not adequate. Though in Nima [1], the humanoid robots have different communication abilities, it could play a doctor, a cook, a chemotherapy hero among others. With each role-playing covering a range of clinical objectives.

Another major finding across the works of literature is the differences in the appearance of the conversational agent. Some works of literature have implemented cartoon character (Robertson) a dog [7], human-sized humanoids [27], humanoids robots [1, 22]. Evaluation of the studies does not provide any explanation for this variety of implementation. This is surprising as the physical appearance of a conversational agent can affect the users' trust and understanding of the system. None of the studies explain in-depth the agents' appearance, such as the age, gender, race or even the realism of the agent. A study by Baylor et al. [3], found that Caucasian students demonstrated a higher level of interest in learning from an African American looking training agent as compared to a Caucasian agent. The spike in interest is largely attributed to the challenge of their expectations as to what a domain expert should look like.

Lastly, while most studies evaluated made use of artificial intelligence algorithms, in some papers [1, 22] the Wizard of Oz technique was applied. In the wizard of oz technique, the wizard is allowed to choose among a given set of replies. Certain times the user is unaware and thinks that interaction is occurring between them and a computer. In Nima [1], a human operator sent commands from a laptop in response to the children. The adoption of oz technique is simpler, it is easier than mapping out possible responses. However, it constrains the interactions and thus is unnatural.

### Future directions

Some notable blank spots exist from the discussion above. To my knowledge, no real-world applications of context-aware voice conversational agents exist for use in the clinical setting. For future works, designing systems for use that can change the tone, pitch and intensity of sound based on interaction with users need to be considered to address context-aware computing in voice user interfaces. Although, this is not without challenges, such as noise from the environment or differing speaking personalities amongst users.

Also, in the aesthetic design of the conversational agent appearance. To account for preferences among users, the option to choose the appearance of the agent should be considered when designing virtual agents. For example, an older user might not take advice from a boy or an animal avatar. Although, this option is not without challenges. Granting users such liberty can potentially diminish the trust between the agent and user if the user feels the agent can appear in different faces. Some users might find this daunting.

## CONCLUSION

This review discusses different aspects of voice user interfaces in clinical settings. Ten papers were reviewed, all of them focused on application in clinical settings. Some in similar use cases and others not. These interfaces have been characterized into input and output modality, use case, dialogue management, and delivery modality. While there exist many reviews of the applications of voice interfaces or conversational agents in healthcare, focus on clinical settings is rare. Finally, this study identified several challenges in the design of current conversational agents and proposed design considerations. Future research would strive to review these interfaces based on task orientation to analyze context; this was not done in this review.

| Citation | Use Case | Delivery Modality | Voice Generator | Nonverbal Cues/ Behavior from System | Nonverbal Cues/ Behavior from users | Dialogue Management (Finite/Frame) | Dialogue Initiative | Input Modality | Output Modality |
| --- | --- | --- | --- | --- | --- | --- | --- | --- | --- |
| *Bickmore, T. W., Laura , P. M., & Brian, J. W. (2009)* | Patient education | Kiosk (tablet) | Voice Synthesizer | Yes | No | Finite | System | Touch | Speech |
| *Pierre, P., Stéphanie, B., Alain, S., Cyril, C., & Jérôme, O. (2015)* | Epworth Sleepiness Scale interviews | 3D animation(gaming computer) | Voice Synthesizer | Yes | No | Finite | System | Speech | Speech |
| *Devault D, A. R. et al. (2014)* | Patient interview (SimSensei) | 3D animation | Prerecorded audio | Yes | Yes | Frame | System | Speech | Speech |
| *Martin B., Fox J. (2006)* | Clinical decision support for cancer patients | Computer (tablet) | | No | No | Complex structure (Frame and Finite) | System | Speech | Speech |
| *Sansen et al. (2016)* | Assistant for aging population and dependent persons | Human-sized humanoid | Voice Synthesizer | Yes | Yes | Frame | Mixed | Speech | Speech |
| *Alemi, M., Meghdari A., Ghanbarzadeh, A. (2014)* | Therapy assistant for child cancer patients (Nima) | Humanoid robot | Voice Synthesizer | Yes | No | Finite | System | Speech | Speech |
| *Meghdari A., Alemi M., Sharaiti, A., Vossoughi, G.R. (2018)* | A buddy for child cancer patients( Arash) | Humanoid robot | Prerecorded audio | Yes | Yes | Finite | System | Speech | Speech |
| *Robertson, S. (2015)* | Shared decision making for prostate cancer patients (eCoach) | 2D animation (Personal/ Desktop computer) | Recorded voice-over audio clips | Yes | No | Finite | System | Speech | Touch |
| *Bott et al. (2019)* | Care coach for older adults | Animated Avatar (tablet) | Voice Synthesizer | Yes | Yes | Frame | Mixed | Speech | Speech |
| *Rizzo, A., Kenny, P., & Parsons, T. (2011)* | Virtual patients for clinical training | PC-based Virtual Reality (VR) systems | Recorded voice | Yes | No | Finite | User | Speech | Speech |

Figure 3: showing a summary of results from the ten works of literature reviewed.

**REFERENCES**

1. Alemi, M., Meghdari, A., Ghanbarzadeh, A., Moghadam, L. J., & Ghanbarzadeh, A. (2014). Impact of a social humanoid robot as a therapy assistant in children cancer treatment. *International Conference on Social Robotics*, *8755*, pp. 11-22. Sydney, NSW, Australia. doi:10.1007/978-3-319-11973-1

2. Al-Shayea, Q. (2011). Artificial Neural Networks in Medical Diagnosis. *International Journal of Computer Science Issues, 8*(2), 150-154.

3. Baylor, A. L. (2009). Promoting motivation with virtual agents and avatars: role of visual presence and appearance. *Philosophical Transactions of the Royal Society B: Biological Sciences*, *364*(1535), 3559–3565. doi: 10.1098/rstb.2009.0148

4. Beveridge, M., & Fox, J. (2006). Automatic generation of spoken dialogue from medical plansand ontologies. *Journal of Biomedical Informatics, 39*, 482-499. doi:0.1016/j.jbi.2005.12.008*Corresponding author. Fax: +44 20 7269 3186.E-mail address:martinbeveridge@slingshot.co.nz(M. Beveridge).

5. Bickmore, T. W., Laura , P. M., & Brian, J. W. (2009). Taking the time to care: empowering low health literacy hospital patients with virtual nurse agents. *CHI '09 Proceedings of the SIGCHI Conference on Human Factors in Computing Systems* (pp. 1265-1274). Boston, MA, USA: ACM.

6. Bosse, D., & Formolo, T. (September 2017). Towards Interactive Agents that Infer Emotions from Voice and Context Information. *EAI Endorsed Transactions on Creative Technologies*. doi:doi: 10.4108/eai.4-9-2017.153054

7. Bott, N., Wexler, S., Drury, L., Pollak, C., Wang, V., Scher, K., & & Narducci, S. (2019). A Protocol-Driven, Bedside Digital Conversational Agent to Support Nurse Teams and Mitigate Risks of Hospitalization in Older Adults: Case Control Pre-Post Study. *Journal of Medical Internet Research, 21*(10). doi:10.2196/13440

8. Brownstein , J., Lannon , J., & Lindenauer, S. (2018). *37 Startups building voice applications for healthcare*. Retrieved from Mobi Health News: mobihealthnews.com/content/37-startups-building-voice-applications-healthcare

9. Cassano, C. (October 01, 2014). The Right Balance – Technology and Patient Care. *Online Journal of Nursing Informatics, 18*(3). Retrieved from https://www.himss.org/right-balance-technology-and-patient-care

10. Cohen , M. H., Giangola, P. J., & Balogh, J. (2004). *Voice User Interface Design.* New Jersey: Addison-Wesley Professional.

11. Devault, D., Artstein, R., Benn, G., T., D., Fast, E., & al., G. e. (2014). SimSensei Kiosk: a virtual human interviewer for healthcare. *International Conference on Autonomous Agents and Multi-agent Systems*, (pp. 1061-1068.). Paris, France.

12. Dourish, P. (2001). Seeking a Foundation for Context-Aware Computing. *Human–Computer Interaction*, *16*(2-4), 229–241. doi: 10.1207/s15327051hci16234_07

13. Jofish Kaye and Paul Dourish. 2014. Special issue on science fiction and ubiquitous computing. *Personal Ubiquitous Comput*. 18, 4 (April 2014), 765-766. http://dx.doi.org/10.1007/s00779-014-0773-4

14. Fadhil, A. (2018). Beyond Patient Monitoring: Conversational Agents Role in Telemedicine and Healthcare Support for Home-Living Elderly Indivduals. doi:arXivpreprint arXiv:1803.06000

15. Gheiratmand M, Rish I, Cecchi GA, Brown MRG, Greiner R, & al., P. P. (2017, May 16). Learning stable and predictive network-based patterns of schizophrenia and its clinical symptoms. *Nature Partner Journal Schizophrenia, 3*(1), 22.

16. Jiang F., Jiang Y., Zhi H., Dong Y., Li H., & al., M. e. (2017, December). Artificial intelligence in healthcare: past, present and future. *Stroke and Vascular Neurology, 2*(4), 230-243

17. Kenny, P., Parsons, T. D., Gratch, J., Leuski, A., & Rizzo, A. A. (n.d.). Virtual Patients for Clinical Therapist Skills Training. *Intelligent Virtual Agents Lecture Notes in Computer Science*, 197–210. doi: 10.1007/978-3-540-74997-4_19

18. Kok J., Boers J.W.E., W.A., K., & P, P. (2009). Artficial Intelligence: Defintion, Trends and Cases. *Artificial intelligence*.

19. Lindsay R., Buchanan B., Feigenbaum E., & J., L. (1980). *Applications of Artificial Intelligence for Organic Chemistry: The DENDRAL Project.* New York, NY: McGraw-Hill.

20. Loh, E. (2018, June 01). Medicine and the rise of the robots: a qualitative review of recent advances of artificial intelligence in health. *Leader, 2*(2), 59-63.

21. Mcdonnell, R., Breidt, M., & Bülthoff, H. H. (2012). Render me real? *ACM Transactions on Graphics*, *31*(4), 1–11. doi: 10.1145/2185520.2335442

22. Meghdari A, Shariati A, Alemi M, Vossoughi GR, Eydi A, Ahmadi E, . . . R., T. (2018, June). Arash: A social robot buddy to support children with cancer in a hospital environment. *Journal of Engineering in Medicine, 232*(6). doi:10.1177/0954411918777520

23. Pierre , P., Stephanie, B., Alain , S., Cyril , C., & Jérôme, O. (2015, March). Could a Virtual Human Be Used to Explore Excessive Daytime Sleepiness in Patients? *Presence Teleoperators & Virtual Environments, 23*(4), 369-376. doi:10.1162/PRES_a_00197

24. Porcheron M., Joel E.F., Reeves S., & S., &. S. (2018). Voice Interfaces in Everyday Life. *CHI '18 Proceedings of the 2018 CHI Conference on Human Factors in Computing Systems.* New York, NY, USA: ACM. doi:https://doi.org/10.1145/3173574.3174214

25. Reenita, D. (March,2 016). Five Technologies That Will Disrupt Healthcare By 2020. Forbes Media Inc. Retrieved from https://www.forbes.com/sites/reenitadas/2016/03/30/top-5-technologies-disrupting-healthcare-by-2020/#553eb7176826

26. Robertson, S., Solomon, R., Riedl, M., Gillespie, T. W., Chociemski, T., & Master. (n.d.). The Visual Design and Implementation of an Embodied Conversational Agent in a Shared Decision-Making Context (eCoach). *Learning and Collaboration Technologies*, 427-437. doi:https://doi.org/10.1007/978-3-319-20609-7_40

27. Sansen, H., Chollet, G., Glackin, C., Jokinen, K., Badii, A., Torres, M. I., . . . Schlögl, S. (2016). The Roberta IRONSIDE project : a cognitive and physical robot coach for dependent persons. *Modelling, Measurement & Control. C: Energetics, Chemistry, Earth, Environmental & Biomedical Problems, 77*(2), 169-181.

28. Son Y., Kim H, .., Kim E., Choi S., & S., L. (2010, Dec). Application of support vector machine for prediction of medication adherence in heart failure patients. *Healthc Inform Res, 16*(4), 253-259.

29. Soronen, H., Pakarinen, S., Hansen, M., Turunen, M., Hakulinen, J., Hella, J., . . . Laivo, T. (2009). User Experience of Speech Controlled Media Center for Physically Disabled Users. *13th International MindTrek Conference: Everyday Life in the Ubiquitous Era*, (pp. 2-5). Tampere, Finland. doi:http://doi.org/10.1145/1621841.1621843

30. Stowers, K. & Mouloua, M. (June 29, 2018). Human Computer Interaction Trends in Healthcare: An Update. *Proceedings of the 2018 International Symposium on Human Factors and Ergonomics in Health Care*, *7 (1)*, pp. 88–91. doi:https://doi.org/10.1177%2F2327857918071019

31. Wang, C., Bickmore, T., Bowen, D. J., Norkunas, T., Campion, M., Cabral, H., . . . Paasche-Orlow, M. (2015, January). Acceptability and feasibility of a virtual counselor (VICKY) to collect family health histories. *Genetics in Medicine, 17*(10), 822–830. doi:10.1038/gim.2014.198

32. Weinstein, C. J. (October 1995). Military and government applications of human-machine. *Human-Machine Communication by Voice. 92*, pp. 10011-10016. Irvine, Ca: National Academy of Sciences at The Arnold and Mabel Beckman Center.

33. Zeki T., Malakooti M., Ataeipoor Y., & S., T. (2012). An Expert System for Diabetes Diagnosis. *American Academic & Scholarly Research Journal, 4*(5), 1-13.